\documentclass[12pt]{article}

\begin{document}
\begin{center}
 {\bf{\Large Black Hole in a Model with Dilaton and Monopole Fields}}
 \vspace{1cm}

E. Kyriakopoulos\footnote{E-mail: kyriakop@central.ntua.gr}\\
Department of Physics\\
National Technical University\\
157 80 Zografou, Athens, GREECE
\end{center}

\begin {abstract}
We present an exact black hole solution in a model having besides
gravity a dilaton and a monopole field. The solution has three
free parameters, one of which can be identified with the monopole
charge, and another with the ADM mass . The metric is
asymptotically flat and has two horizons and irremovable
singularity only at $r=0$. The dilaton field is singular only at
$r=0$. The dominant and the strong energy condition are satisfied
outside and on the external horizon. According to a formulation of
the no hair conjecture the solution is "hairy". Also the well know
GHS-GM solution is obtained from our solution for certain values
of its parameters.

PACS number(s): 04.20.Jb, 04.70.Bw, 04.20.Dw

\end {abstract}

 There are various formulations of the no hair conjecture
[1-5]. According to the original one \cite{Wh} a black hole is
uniquely determined by its mass its electric and magnetic charges
and its angular momentum. There are no other conserved quantities
 provided that they cannot be expressed in terms of the above
quantities. This happens in the well known solution of D.
Garfinkle, G. T. Horowitz and A. Strominger (GHS) \cite{Ga}, found
previously by G. W. Gibbons and by G. W. Gibbons and K. Maeda
\cite{Gi}, which has two free parameters namely the
Arnowitt-Deser-Misner (ADM) mass and the magnetic charge and in
addition one more parameter the dilaton charge, which however can
be expressed in terms of the other two. Therefore the GHS solution
does not violate the no hair theorem in its original formulation
\cite{Ho,Su}.
 Another formulation of the no hair
conjecture is the following \cite{Nu}:"We say that in a given
theory there is black hole hair when the space-time metric and the
configuration of the other fields of a stationary black hole
solution are not completely specified by the conserved charges
defined at asymptotic infinity".

In this work we present, in a model having besides gravity a
dilaton and a monopole field, an exact solution which is
characterized by three independent quantities namely the monopole
charge the ADM mass and another free parameter. The solution is
simple and has two horizons. The scalar field is singular only at
the origin and the metric has an essential singularity only at the
origin. The metric is written in Eddington-Finkelstein type
coordinates. Also one coordinate patch of Kruskal type coordinates
is given. According to Ref. \cite{Nu} our solution is "hairy".

Consider the action
\begin{equation}
\int d^{4}x \sqrt{-g} L = \int d^{4}x \sqrt{-g}
\{R-\frac{1}{2}\partial_{\mu}\psi\partial^{\mu}\psi-
(g_{1}e^{\psi} +g_{2}e^{-\psi})F_{\mu\nu}F^{\mu\nu}\} \label{1}
\end{equation}
where $R$ is the Ricci scalar, $\psi$ is a dilaton field and
$F_{\mu\nu}$ is a pure monopole field
\begin{eqnarray}F=Q sin\theta d\theta\wedge d\phi \label{2}
\end{eqnarray}
with $Q$ its magnetic charge. From this action we find the
following equations of motion
\begin{equation}
(\partial^{\rho}\psi)_{;\rho}-(g_{1}e^{\psi}-
g_{2}e^{-\psi})F_{\mu\nu}F^{\mu\nu}=0 \label{4}\end{equation}
\begin{equation}
((g_{1}e^{\psi}+g_{2}e^{-\psi})F^{\mu\nu})_{;\mu}=0 \label{5}
\end{equation}
\begin{equation}R_{\mu\nu}=\frac{1}{2}\partial_{\mu}\psi\partial_{\nu}\psi+
2(g_{1}e^{\psi}+g_{2}e^{-\psi})(F_{\mu\sigma}{F_{\nu}}^{\sigma}-
\frac{1}{4}g_{\mu\nu}F_{\rho\sigma}F^{\rho\sigma})\label{6}
\end{equation}

We want to find static spherically symmetric solutions of the
above equations which are asymptotically flat and have regular
horizon. We write the metric in the form \cite{Ga}
\begin{equation}
ds^2=-\lambda^{2}dt^{2}+\lambda^{-2}dr^{2}+\xi^{2}d\Omega
\label{7}
\end{equation}
where $\lambda$ and $\xi$ are functions of $r$ only and
$d\Omega=d\theta^{2}+sin^{2}\theta {d{\phi}^{2}}$. From the above
metric and Eq (\ref{2}) we get
$F_{\mu\nu}F^{\mu\nu}=\frac{2Q^{2}}{\xi^{4}}$, and we can prove
that Eq (\ref{5}) is satisfied. The dilaton Eq. (\ref{4}) takes
the form
\begin{equation}
(\lambda^{2}\xi^{2}\psi')'=2(g_{1}e^{\psi}-
g_{2}e^{-\psi})Q^{2}\xi^{-2}\label{8}
\end{equation}
where prime denotes differentiation with respect to $r$. The
non-vanishing components of the Ricci tensor of the metric
(\ref{7}) are $R_{00}$, $R_{11}$, $R_{22}$ and
$R_{33}=sin^{2}\theta{R_{22}}$, and for the first three components
we get respectively from Eqs (\ref{6}) the relations
\begin{equation}
(\lambda^{2})'' +(\lambda^{2})'(\xi^{2})'\xi^{-2}=2(g_{1}e^{\psi}+
g_{2}e^{-\psi})Q^{2}\xi^{-4} \label{9}
\end{equation}
\[-(\lambda^{2})''\lambda^{-2}-2(\xi^{2})''\xi^{-2}-
(\lambda^{2})'(\xi^{2})'\lambda^{-2}\xi^{-2}+
[(\xi^{2})']^{2}\xi^{-4}=(\psi')^{2}\]
\begin{equation}-2(g_{1}e^{\psi}+
g_{2}e^{-\psi})Q^{2}\lambda^{-2}\xi^{-4} \label{10 }
\end{equation}
\begin{equation}
-[\lambda^{2}(\xi^{2})']'+2=2(g_{1}e^{\psi}+
g_{2}e^{-\psi})Q^{2}\xi^{-2} \label{11}
\end{equation}

Eqs (\ref{8})-(\ref{11}) form a system of four equations for the
three unknowns $\lambda^{2}$, $\xi^{2}$ and $\psi$. We found the
following solution of this system
\begin{equation}
\lambda^{2}=\frac{(r+A)(r+B)}{r(r+\alpha)},\>\>\>\>\xi^{2}=r(r+\alpha)\label{12}
\end{equation}
\begin{equation}
e^{\psi}=e^{\psi_{0}}(1+\frac{\alpha}{r})  \label{13}
\end{equation}
where $A$, $B$, $\alpha$ and $\psi_{0}$ are integration constants,
provided that the following relations are satisfied
\begin{equation}
g_{1}=\frac{AB}{2Q^{2}}e^{-\psi_{0}},
\>\>\>\>\>g_{2}=\frac{(\alpha-A)(\alpha-B)}{2Q^{2}}e^{\psi_{0}}
\label{14}
\end{equation}
From Eqs (\ref{7}) and (\ref{12}) we get
\begin{equation}
ds^{2}=-\frac{(r+A)(r+B)}{r(r+\alpha)}dt^{2}+\frac{r(r+\alpha)}{(r+A)(r+B)}dr^{2}+
r(r+\alpha)d{\Omega}  \label{15}
\end{equation}
Our solution is given by Eqs (\ref{2}) and (\ref{13})-(\ref{15}).

From Eq. (\ref{15}) we get asymptotically
$-g_{00}=1-\frac{\alpha-A-B}{r} + O(r^{-2})$.
 Therefore the solution
is asymptotically flat and its ADM mass $M$ is given by
\begin{equation}
2M=\alpha-A-B \label{17}
\end{equation}

It is obvious that $\psi_{0}$ is the asymptotic value of $\psi$.
Also for the choice
\begin{equation}
\alpha>0,\>\>\>\>\> A<0,\>\>\>\>\> B<0 ,\label{19}
\end{equation}
we have
\begin{equation}
g_{1}>0,\>\>\>\>\> g_{2}>0,\>\>\>\>\> M>0,\label{20}
\end{equation}
and $\psi$ is singular only at $r=0$. We shall make this choice
for $\alpha$, $A$ and $B$.

The solution has the integration constants $\alpha$, $A$, $B$,
$\psi_{0}$ and $Q$, which for given $g_{1}$ and $g_{2}$ must
satisfy Eqs (\ref{14}). Therefore only three of them are
independent. Introducing the ADM mass $M$ by the relation
(\ref{17}) we can take $M$, $Q$ and $\psi_{0}$ as independent
parameters and express the parameters $\alpha$, $A$ and $B$ of the
dilaton field and the metric in terms of then. This means that the
solution has arbitrary mass, arbitrary magnetic charge and an
additional arbitrary parameter. Therefore it is a "hairy" solution
according to the definition of such a solution given in Ref.
\cite{Nu}.

The metric coefficient $g_{rr}$ is singular at $r=-A$ and  $r=-B$
while the coefficient $g_{tt}$ is singular at $r=0$ but not at
$r=-\alpha$, since $\alpha>o$. However the singularities at $r=-A$
and $r=-B$ are coordinate singularities. Indeed the Ricci scalar
$R$ and the curvature scalar
$R_{\mu\nu\rho\sigma}R^{\mu\nu\rho\sigma}$ are given by
\begin{equation}
R=\frac{\alpha^{2}(r+A)(r+B)}{2r^{3}(r+\alpha)^{3}},\>\>\>\>\>
R_{\mu\nu\rho\sigma}R^{\mu\nu\rho\sigma}=
\frac{P(r,\alpha,A,B)}{4r^{6}(r+\alpha)^{6}} \label{21}
\end{equation}
where $P(r,\alpha,A,B)$ is a complicated polynomial of $r$,
$\alpha$ $A$ and $B$. Therefore $R$ and
$R_{\mu\nu\rho\sigma}R^{\mu\nu\rho\sigma}$ are regular at $r=-A$
and $r=-B$ , which means that only at $r=0$ we have a real, an
irremovable  singularity.

If we make the Eddington-Finkelstein type transformation
\begin{equation}
t=t' \pm \{\alpha lnr + \frac{(\alpha-A)A}{B-A}ln|r+A| +
\frac{(\alpha-B)B}{A-B}ln|r+B|\},\>\>r=r',\>\>\theta={\theta}',\>\>
\phi={\phi}'\label{22}
\end{equation}
the metric $ds^{2}$ of Eq. (\ref{15}) takes the regular at $r=-A$
and $r=-B$ form
\[ds^{2}=-\frac{(r'+A)(r'+B)}{r'(r'+\alpha)}dt'^{2}+\frac{r'+\alpha}{r'^{3}}\{r'^{2}-(A+B)r'-
AB\}dr'^{2}\]
\begin{equation}
 \mp \frac{2}{r'^{2}}\{(A+B)r'+AB\}dr'dt' +
r'(r'+\alpha)d\Omega \label{23 }
\end{equation}
From the above expression we find that the radial null directions,
i.e. the directions for which $ds^{2}=d \theta=d \phi=0$, are
determined by the relations
\begin{equation}
r'dt'\pm(r'+\alpha)dr'=0 \label{24}
\end{equation}
\begin{equation}
r'(r'+A)(r'+B)dt' \mp (r'+\alpha)\{r'^{2}-(A+B)r'-AB\}dr'=0
\label{25}
\end{equation}
Integrating the above relations we get
\begin{equation}
t'=\mp r' \mp \alpha lnr' + const \label{26}
\end{equation}
\begin{equation}
t'=\pm r' \mp \alpha lnr' \pm \frac{2A(\alpha-A)}{A-B}ln|r'+A| \pm
\frac{2B(\alpha-B)}{B-A}ln|r'+B| + const \label{27}
\end{equation}
which are the equations of the intersections of the light cone
with the $t'-r'$ plane.

Proceeding as in the Eddington-Finkelstein treatment of the
Schwarzshild solution and the Reissner-Nordstrom solution we can
draw a space-time diagram and show that the solution with the
upper sign is a black hole solution. The solution has two horizons
at $r'=r=-A$ and at $r'=r=-B$. Assume that $|A| > |B|$ and call
$I$, $II$ and $III$ the regions with $r>-A$, $-A>r>-B$ and
$-B>r>0$ respectively. Then we can show that particles from region
$I$ can cross the surface $r=-A$ and enter region $II$, while a
particle in region $II$ will reach the surface $r=-B$
asymptotically or it will cross it and enter region $III$. A
particle cannot move from region $III$ to region $II$. The
solution with the lower sign is a white hole solution.

Since we have two horizons we need two Kruskal type coordinate
patches to cover the whole range of $r$, as in the Reissner-
Nordstrom case  \cite{Gr}. To introduce the first patch assume as
before that $-A>-B$, let $-A>r_c>-B$ and define new coordinates
 $u$ and $v$ for $r>-A$ by the
relations
\begin{equation}
u=\frac{2A(\alpha-A)}{A-B}e^{\frac{(A-B)r}{2A(\alpha-A)}}(r+A)^\frac{1}{2}
(r+B)^{-\frac{B(\alpha-B)}{2A(\alpha-A)}}cosh\{\frac{(A-B)t}{2A(\alpha-A)}\}
\label{28}
\end{equation}
\begin{equation}
v=\frac{2A(\alpha-A)}{A-B}e^{\frac{(A-B)r}{2A(\alpha-A)}}(r+A)^\frac{1}{2}
(r+B)^{-\frac{B(\alpha-B)}{2A(\alpha-A)}}sinh\{\frac{(A-B)t}{2A(\alpha-A)}\}
\label{29}
\end{equation}
and for $-A>r>r_c$ by the relations
\begin{equation}
u=\frac{2A(\alpha-A)}{A-B}e^{\frac{(A-B)r}{2A(\alpha-A)}}|r+A|^\frac{1}{2}
(r+B)^{-\frac{B(\alpha-B)}{2A(\alpha-A)}}sinh\{\frac{(A-B)t}{2A(\alpha-A)}\}
\label{30}
\end{equation}
\begin{equation}
v=\frac{2A(\alpha-A)}{A-B}e^{\frac{(A-B)r}{2A(\alpha-A)}}|r+A|^\frac{1}{2}
(r+B)^{-\frac{B(\alpha-B)}{2A(\alpha-A)}}cosh\{\frac{(A-B)t}{2A(\alpha-A)}\}
\label{31}
\end{equation}
Then Eq. (\ref{15}) for $r>r_c$ takes the form
\begin{equation}
ds^{2}=\frac{1}{r(r+\alpha)}(r+B)^{1+\frac{B(\alpha-B)}{A(\alpha-A)}}
e^{\frac{(B-A)r}{A(\alpha-A)}}(du^{2}-dv^{2})+r(r+\alpha)d{\Omega}
\label{32}
\end{equation}
where, if we know $u$ and $v$, $r$ and $t$ are determined for
$r>-A$ by the relations
\begin{equation}
u^{2}-v^{2}=\{\frac{2A(\alpha-A)}{A-B}\}^{2}e^\frac{(A-B)r}{A(\alpha-A)}
(r+A)(r+B)^{-\frac{B(\alpha-B)}{A(\alpha-A)}} \label{33}
\end{equation}
\begin{equation}
\frac{v}{u}=tanh\{\frac{(A-B)t}{2A(\alpha-A)}\} \label{34}
\end{equation}
and for $-A>r>r_c$ by analogous relations. The factor $r+A$ of Eq.
(\ref{15}) has been removed in Eq. (\ref{32}). Introducing an
analogous second coordinate patch we can remove the factor $r+B$.
Of course we cannot eliminate both factors simultaneously.

The energy-momentum tensor $T_{\mu\nu}$ of our solution is given
by
\[T_{\mu\nu}=\partial_{\mu}\psi\partial_{\nu}\psi
+4(g_{1}e^{\psi}+ g_{2}e^{-\psi})F_{\mu\rho}{F_{\nu}}^{\rho}
-g_{\mu \nu}\{\frac{1}{2}\partial_{\rho}\psi\partial^{\rho}\psi
+(g_{1}e^{\psi}\]
\[+g_{2}e^{-\psi})F_{\rho \sigma}F^{\rho \sigma} \}
=\frac{\alpha^{2}}{r^{2}(r+\alpha)^{2}}\delta_{\mu r}\delta_{\nu
r}+\{\frac{2AB}{r^{2}}+\frac{2(\alpha-A)(\alpha-B)}{(r+\alpha)^{2}}\}
(\delta_{\mu\theta}\delta_{\nu \theta} \]
\begin{equation}
+{sin^{2}\theta}\delta_{\mu \phi}\delta_{\nu \phi}) -g_{\mu \nu}\{
\frac{{\alpha^{2}}(r+A)(r+B)}{2r^{3}(r+\alpha)^{3}}+\frac{AB}{r^{3}(r+\alpha)}
 +\frac{(\alpha-A)(\alpha-B)}{r(r+\alpha)^{3}}\}
\label{35}
\end{equation}
Calculating the eigenvalues of $T_{\mu\nu}$ we can show that it
satisfies the dominant as well as the strong energy condition
outside and on the external horizon.

If $-A>-B>0$ and $\alpha>0$ the Hawking temperature $T_H$ of our
solution is \cite{Ho}
\begin{equation}
T_H=\frac{|(\lambda^{2})'(-A)|}{4\pi}=\frac{B-A}{4\pi A(A-\alpha)}
\label{36}
\end{equation}

Assume that
\begin{equation}
\alpha=B=\frac{2Q^2}{A}e^{\psi_0} \label{37}
\end{equation}
Then from Eqs (\ref{14}) and (\ref{17}) we get
\begin{equation}
 g_1=1,\>\>\>\> g_2=0,\>\>\>\> 2M=-A \label{38}
\end{equation}
and Eqs (\ref{13}) and (\ref{15}) become
\begin{equation}
 e^\psi=e^{\psi_0}(1-\frac{Q^2}{Mr}e^{\psi_0}) \label{39}
\end{equation}
\begin{equation}
 ds^2=-(1-\frac{2M}{r})dt^2+(1-\frac{2M}{r})^{-1}dr^2
 +r(r-\frac{Q^2}{M}e^{\psi_0})d\Omega\label{40}
\end{equation}
If in Eqs (\ref{39}) and (\ref{40}) we make the replacements
\begin{equation}
 \psi\rightarrow-2\phi, \>\>\>\>\psi_0\rightarrow-2\phi_0 \label{41}
\end{equation}
we get the GHS-GM solution. Therefore the GHS-GM solution is a
special case of our solution.

 I am very grateful to A. Kehagias for
many illuminating discussions and suggestions. The computation of
the Ricci scalar and the curvature scalar was done with the help
of a programm given to me by S. Bonanos, whom I thank.

\end{document}